\documentclass[lettersize,journal]{IEEEtran}
\usepackage{amsmath,amsfonts}
\usepackage{algorithmic}
\usepackage{array}
\usepackage[caption=false,font=normalsize,labelfont=sf,textfont=sf]{subfig}
\usepackage{textcomp}
\usepackage{stfloats}
\usepackage{multirow}
\usepackage{booktabs}
\usepackage{algorithm}
\usepackage{algorithmic}
\usepackage{url}
\usepackage{verbatim}
\usepackage{graphicx}
\def\BibTeX{{\rm B\kern-.05em{\sc i\kern-.025em b}\kern-.08em
    T\kern-.1667em\lower.7ex\hbox{E}\kern-.125emX}}
\usepackage{balance}
\def\method{DDGF}
\begin{document}
\title{Diffusion-Driven Generative Framework for Molecular Conformation Prediction}
\author{Bobin Yang, Jie Deng, Zhenghan Chen, Ruoxue Wu
\thanks{Bobin Yang is with the Life Sciences program at the University of Chinese Academy of Sciences. (e-mail: yangbolin22@mails.ucas.ac.cn) }
\thanks{Jie Deng is with the B.E. in Software Engineering at Nankai University. (e-mail: dengjie2024@126.com) }
\thanks{Zhenghan Chen is with the applied scientist in Yanshi. (e-mail: pandaarych@gmail.com) }
\thanks{Ruoxue Wu is with the the M.S. degree in software engineering from Yunnan University. (e-mail: rochelle.wu820@gmail.com) }
\thanks{Bobin Yang and Jie Deng contributed equally}
\thanks{Corresponding author: Zhenghan Chen}
}

\markboth{Journal of \LaTeX\ Class Files,~Vol.~18, No.~9, September~2020}%
{How to Use the IEEEtran \LaTeX \ Templates}

\maketitle

\begin{abstract}
The task of deducing three-dimensional molecular configurations from their two-dimensional graph representations holds paramount importance in the fields of computational chemistry and pharmaceutical development. This process significantly contributes to our comprehension of molecular mechanisms and interactions. The rapid advancement of machine learning, particularly within the domain of deep generative networks, has revolutionized the precision of predictive modeling in this context.
Traditional approaches often adopt a two-step strategy: initially estimating interatomic distances and subsequently refining the spatial molecular structure by solving a distance geometry problem. However, this sequential approach occasionally falls short in accurately capturing the intricacies of local atomic arrangements, thereby compromising the fidelity of the resulting structural models.
Addressing these limitations, this research introduces a cutting-edge generative framework named \method{}. This framework is grounded in the principles of diffusion observed in classical non-equilibrium thermodynamics. \method{} views atoms as discrete entities and excels in guiding the reversal of diffusion, transforming a distribution of stochastic noise back into coherent molecular structures through a process akin to a Markov chain. This transformation commences with the initial representation of a molecular graph in an abstract latent space, culminating in the realization of three-dimensional structures via a sophisticated bilevel optimization scheme meticulously tailored to meet the specific requirements of the task.
One of the formidable challenges in this modeling endeavor involves preserving roto-translational invariance to ensure that the generated molecular conformations adhere to the laws of physics. Our proposed framework adeptly navigates this complexity, facilitating comprehensive training from inception to conclusion through the fine-tuning of a weighted variational lower bound that considers the conditional probability of conformations.
Extensive experimental evaluations confirm the efficacy of the proposed \method{} in comparison to state-of-the-art methods.
\end{abstract}

\section{Introduction}

Graph representation learning has achieved remarkable success in modeling molecules for various tasks, encompassing property prediction~\cite{yin2022deal,duvenaud2015convolutional}, graph domain adaptation~\cite{anonymous2024dream,yin2023,yin2023coco}, time series prediction~\cite{yin2022dynamic,yin2023messages}, and molecule generation~\cite{Jin_Barzilay_Jaakkola_2017,shi2020graphaf}. In this approach, molecules are typically represented as atom-bond graphs. Although this representation proves effective in multiple applications, a more intrinsic and informative depiction of molecules resides in their 3D geometry, often referred to as conformation. In this conformation, atoms are depicted using their Cartesian coordinates. The 3D structures exert significant influence over the biological and physical characteristics of molecules, rendering them crucial in various domains such as computational drug and material design ~\cite{thomas2018tensor,gebauer2022inverse,jing2021learning,batzner20223}.
Despite these advancements, predicting stable molecular conformations remains a formidable challenge. Conventional approaches reliant on molecular dynamics (MD) or Markov chain Monte Carlo (MCMC) methods are computationally intensive and struggle to efficiently handle large molecules~\cite{Hawkins_2017}. To address this challenge, researchers are exploring novel methods that integrate graph-based models with physics-based simulations or leverage reinforcement learning techniques to optimize molecular conformations.

Experimentally determining the three-dimensional arrangements of molecules can be a challenging and resource-intensive endeavor, characterized by high costs and lengthy timeframes. As a result, the pursuit of computationally deriving precise and energetically favorable molecular structures has gained prominence within the field of computational chemistry. Traditional approaches in this domain have traditionally relied on methods such as Markov chain Monte Carlo (MCMC) or molecular dynamics (MD) simulations~\cite{de2016role}. These methods involve proposing potential conformations and assigning energy levels to them through either cost-effective empirical potential models or high-precision quantum chemical calculations ~\cite{ballard2015exploiting}.
Recently, there has been a shift towards the integration of machine learning algorithms to estimate the conditional probability distribution $p(\mathbf{C}|G)$ for generating stable conformations $\mathbf{C}$ of a given molecular graph $G$. This approach builds upon datasets of molecules with known stable conformations ~\cite{mansimov2019molecular,simm2019generative}. Notably, two pioneering studies \cite{simm2019generative,xu2021learning} have introduced a sequential methodology. The initial phase involves predicting interatomic distances, which lays the foundation for the subsequent creation of molecular structures based on these predicted distances. This task is facilitated by solving a set of distance geometry challenges~\cite{Liberti_Lavor_Maculan_Mucherino_2012}. By leveraging the principles of distance geometry, these innovative methods successfully capture the essential characteristics of rotation and translation invariance in molecular structures, leading to increasingly accurate predictions of molecular conformations.

The sequential methodologies that initially estimate interatomic distances and then deduce conformations are susceptible to a notable drawback: the potential for inaccuracies in local three-dimensional geometries. There are instances where the models predicting distances may inadvertently endorse geometrically incompatible sets of distances, leading to significant deviations. These inaccuracies, when fed into the conformation generation stage, can produce atypical and implausible molecular structures.
The source of such discrepancies can often be attributed to the distance prediction models, which prioritize maximizing individual distance likelihoods rather than ensuring the holistic accuracy of the resulting conformations. To address this challenge, our research introduces a streamlined, end-to-end framework that directly formulates conformation predictions from the molecular graph, eliminating the intermediate distance prediction phase. Our approach aims to comprehensively capture the intricate dependencies and interactions among atoms, thus striving to enhance the precision and reliability of conformation prediction.
This shift towards an integrated predictive model aligns with the evolution observed in the field of protein structure prediction. Innovations exemplified by the AlphaFold2 algorithm underscore the value and impact of adopting such an end-to-end approach. AlphaFold2's improvements over its predecessor, AlphaFold, highlight the advantages of consolidating distinct phases into a cohesive modeling process ~\cite{senior2020improved,jumper2020high}. Although AlphaFold2's internal mechanisms are not fully disclosed, its remarkable success in deducing protein structures from linear sequences underscores the potential of end-to-end methodologies in complex structural prediction tasks.

In our research, we introduce \method{}, an advanced and integrated framework specifically designed for the synthesis of molecular conformations using a sophisticated bilevel programming approach. A key aspect of our method is the incorporation of roto-translational invariance within the conformational modeling scheme, where interatomic distances play a pivotal role. Our innovative approach goes beyond traditional distance prediction models that aim to minimize prediction errors in the distance space.
By framing the problem within the context of bilevel optimization~\cite{franceschi2018bilevel,yin2022generic}, \method{} simultaneously addresses both the prediction of distances and the resolution of distance geometry challenges that are fundamental to conformational synthesis. To the best of our knowledge, \method{} is the first algorithm in the field of molecular conformation synthesis that is trainable end-to-end while preserving the crucial aspects of roto-translational invariance.
Our extensive experiments demonstrate the exceptional capabilities of \method{}, consistently outperforming the current leading methods across a range of well-established benchmarks covering conformational synthesis and distance distribution analysis. Furthermore, our findings underscore the importance of an end-to-end training objective in the generation of conformations that are not only realistic but also chemically meaningful.

\section{Related Work}

\textbf{Conformation generation}. 
Recent advancements in computational molecular modeling have given rise to a multitude of innovative deep generative models and methodologies, each presenting its unique advantages and inherent challenges. For instance, the CVGAE \cite{mansimov2019molecular} harnessed a variational autoencoder (VAE) for the direct computation of 3D atomic structures. However, this model faced limitations in adequately preserving the crucial property of roto-translation equivariance, ultimately resulting in suboptimal performance.
In response to this limitation, newer models have turned to intermediate structural descriptors, such as atomic distances and dihedral angles, known for their roto-translationally invariant nature, which is essential for accurately representing molecular shapes\cite{kohler2020equivariant}. Yet, this modular approach, relying on these intermediate descriptors, introduces certain inefficiencies and complexities during both the training phase and sample generation. Approaches like those proposed by \cite{simm2019generative} and \cite{xu2021learning} involve predicting interatomic distance matrices and subsequently applying Distance Geometry solutions to derive spatial coordinates ~\cite{Liberti_Lavor_Maculan_Mucherino_2012}.
Although CONFVAE represented a step forward with its unified, bilevel optimization-based end-to-end framework ~\cite{xu2021end}, these techniques still grapple with the issue of error magnification. Inaccuracies in distance estimations often lead to misguidance of the coordinate determination mechanisms, resulting in the creation of molecular structures that are not only inaccurate but sometimes structurally implausible.

In an attempt to tackle this challenge, the CONFGF model, as introduced in works by~\cite{shi2021learning,luo2021predicting}, aimed to learn the gradient of log-likelihood concerning coordinates. Nevertheless, in practical applications, the model still depended on intermediate geometric elements. It estimated the gradient with respect to interatomic distances using denoising score matching (DSM)~\cite{song2019generative,song2020improved} and subsequently applied the chain rule to derive the gradient of coordinates. Unfortunately, by utilizing DSM (Distance Gradient-based learning), the model was trained using perturbed distance matrices, which had the potential to violate the triangular inequality or contain negative values. Consequently, the model acquired knowledge from these incorrect distance matrices but was evaluated using valid ones computed from coordinates. This discrepancy between training and testing data resulted in a significant out-of-distribution issue ~\cite{hendrycks2016baseline}.
A recent approach called GEOMOL~\cite{ganea2021geomol} introduced a highly systematic rule-based pipeline. It primarily focused on predicting a minimal set of geometric quantities, such as lengths and angles, and then reconstructed both local and global structures of conformations through a sophisticated procedure.
Additionally, there have been efforts to leverage reinforcement learning for conformation search, exemplified by methods like TorsionNet~\cite{TorsionNet}. However, it's important to note that this approach fundamentally differs from other existing methodologies as it relies on the rigid rotor approximation and only models torsion angles.


\textbf{Advances in Protein Topology Prediction}. 
The field of protein topology prediction has experienced rapid expansion, with a multitude of innovative methodologies emerging. These innovations encompass a range of approaches, including the use of flow-based models~\cite{noe2019boltzmann,xu2021end}, which effectively map the conformations of protein main chains. Additionally, techniques employing recurrent neural architectures have been employed for the sequential modeling of amino acids ~\cite{alquraishi2019end}. ~\cite{ingraham2018learning} introduced a novel approach that utilizes neural networks to emulate an energy landscape, facilitating the inference of protein folds. The groundbreaking AlphaFold algorithm~\cite{senior2020improved,jumper2021highly} has significantly raised the bar in terms of accuracy for protein structure prediction.
It is important to highlight, however, that proteins are structurally simple, primarily linear molecules, which is in contrast to the complex and branched configurations often found in general molecules, including various ring structures. This structural disparity makes the direct application of protein-folding methodologies less suitable for the intricate task of predicting the conformation of general molecular structures. Hence, there is a compelling need for distinct modeling techniques that are tailored to address the unique challenges posed by non-protein molecular structures.


\textbf{Point cloud generation}. 
In the domain of three-dimensional structural synthesis, diffusion-driven models have emerged as significant contributors, as evidenced by pioneering efforts in recent literature~\cite{luo2021diffusion,chibane2020implicit}. These approaches have primarily focused on the generation of point clouds, which represents a distinct area of study from our current research. Point clouds typically lack the graph-like structure that encompasses diverse atomic and bonding information, which is a fundamental aspect of our model's domain. Moreover, the principle of equivariance, although not extensively incorporated in these point cloud methodologies, constitutes a cornerstone in our approach. This disparity in foundational concepts underscores the novelty of our model when compared to the landscape of existing methods.

In conclusion, the realm of deep generative models has ushered in a new era for the generation of molecular conformations. Each innovation brings unique advantages while also encountering specific challenges. There remains an ongoing necessity for exploration and advancement to overcome the existing hurdles in improving the accuracy and computational efficiency of these conformational generation techniques.

\section{Preliminaries}
\subsection{Notations and Problem Definition}
\textbf{Notations}. In this paper, we delineate a molecule comprising $n$ atoms as a graph $G=(V, E)$ without directionality. Here, $V={v_i}{i=1}^n$ signifies the vertices corresponding to atoms, and $E = {e{ij} | (i, j) \in {1, \ldots, n}^2 }$ represents the bonds between them as edges. Every vertex $v_i$ encapsulates atomic properties such as atomic number or element category. Interatomic relationships are denoted by edges $e_{ij}$, each annotated with its bond type. Edges without a chemical bond are designated a distinct virtual category. Geometrically, vertices in $V$ are each associated with a 3D coordinate vector $\mathbf{c}_i \in \mathbb{R}^3$, situating the atom in a tridimensional framework. The entirety of these coordinates forms the molecular conformation matrix $\mathbf{C} = [\mathbf{c}_1, \mathbf{c}_2, \ldots , \mathbf{c}_n]$ within $\mathbb{R}^{n \times 3}$ space.

\textbf{Problem Definition}. The task of molecular conformation generation is a conditional generative problem, where we are interested in generating stable conformations for a provided graph $G$. Consider a set of molecular graphs $G$ with each graph's conformation $\mathbf{C}$ acting as independently and identically distributed samples emanating from a latent Boltzmann distribution~\cite{noe2019boltzmann}. The crux of our endeavor lies in devising a generative model denoted by $p_\theta (\mathbf{C}|G)$. This model, which should facilitate straightforward sample generation, is tasked with emulating the probabilistic nature of the Boltzmann distribution.

\subsection{Equivariance}
Equivariance plays a vital role in machine learning models that deal with atomic systems. It is the principle that certain properties of atoms, such as the vectors representing atomic dipoles or forces, should transform consistently with the transformations applied to the atomic coordinates. For example, if a molecule is rotated in space, the direction of the dipoles or forces should rotate in the same way~\cite{thomas2018tensor,weiler20183d,fuchs2020se,miller2020relevance,batzner20223}.
Incorporating this inductive bias directly into the architecture of a machine learning model has been proven to be highly effective for accurately modeling three-dimensional geometries. By ensuring that the model's outputs change predictably with changes to the inputs, we can greatly enhance the model's ability to generalize from one situation to another~\cite{kohler2020equivariant}. This is particularly important in the field of molecular modeling, where the ability to predict how a molecule will behave under various spatial transformations is crucial.
Formally, a function F : X → Y is equivariant w.r.t a group $G$ if:
\begin{equation}
    \mathcal{F}\odot T_g(x)=S_g \odot \mathcal{F}(x),
\end{equation}
where $T_g$ and $S_g$ are transformations for an element $g\in G$, acting on the vector spaces $X$ and $Y$, respectively. 

\subsection{Equivariant Graph Neural Networks}
EGNNs~\cite{satorras2021n} are a type of Graph Neural Network that satisfies the equivariance constraint. In this work, we consider interactions between all atoms, and therefore assume a fully connected graph $G$ with nodes $v_i \in V$. Each node $v_i$ is endowed with coordinates $\mathbf{c}_i \in \mathbb{R}^3$ as well as features $h_i \in \mathbb{R}^d$. In this setting, EGNN consists of the composition of Equivariant Convolutional Layers $\mathbf{c}^{l+1}, \mathbf{h}^{l+1} = EGCL[\mathbf{c}^l, \mathbf{h}^l]$ which are defined as:
\begin{equation}
\label{egnn}
\begin{aligned}
\mathbf{m}_{ij}=\phi_e &(\mathbf{h}_i^l,\mathbf{h}_j^l,d_{ij}^2,a_{ij}),\quad \mathbf{h}_i^{l+1}=\phi_h (\mathbf{h}_i^l,\sum_{j\neq i} \Tilde{e}_{ij}\mathbf{m}_{ij}),\\
\mathbf{c}_i^{l+1}&=\mathbf{c}_i^l+\sum_{j\neq i}\frac{\mathbf{c}_i^l-\mathbf{c}_j^l}{d_{ij}+1}\phi_x(\mathbf{h}_i^l,\mathbf{h}_j^l,d_{ij}^2,a_{ij}),
\end{aligned}
\end{equation}
where $l$ indexes the layer, and $d_{ij} = ||\mathbf{c}^l_i - \mathbf{c}^l_j||_2$ is the euclidean distance between nodes $(v_i, v_j)$, and $a_{ij}$ are optional edge attributes. The difference $(\mathbf{c}^l_i - \mathbf{c}^l_j)$ in Equation \ref{egnn} is normalized by $d_{ij} + 1$ as done in~\cite{garcia2021n} for improved stability, as well as the attention mechanism which infers a soft estimation of the edges $ \Tilde{e}_{ij} = \phi_{inf} (\mathbf{m}_{ij} )$. All learnable components ($\phi_e$,$\phi_h$, $\phi_x$ and $\phi_{inf}$ ) are parametrized by fully connected neural networks. An entire EGNN architecture is then composed of $L$ EGCL layers which applies the following non-linear transformation $\hat{\mathbf{c}}, \hat{\mathbf{h}} = EGNN[\mathbf{c}^0, \mathbf{h}^0]$.

\subsection{Bilevel Optimization}
Bilevel programs are a class of optimization problems in which a group of variables participating in the primary (outer) objective function are determined through the solution of a secondary (inner) optimization problem. To define this formally, we have the outer objective function denoted as F, the inner objective function as H, and the respective outer and inner variables represented as $\theta$ and $w$, a bilevel program can be formulated by:
\begin{equation}
\label{bilevel}
    \min_\theta F(w_\theta)\ such\ that\ w_\theta \in arg \min_w H(w,\theta).
\end{equation}
Bilevel programs have proven to be effective in various scenarios, including hyperparameter optimization, adversarial and multi-task learning, as well as meta-learning.

Typically solving equation~\ref{bilevel} is intractable since the solution sets of $w_\theta$ may not be available in closed form. A common approach is to replace the exact minimizer of the inner object $H$ with an approximation solution, which can be obtained through an iterative optimization dynamics $\Phi$ such as stochastic gradient descent (SGD). Starting from the initial parameter $w_0$, we can get the approximate solution $w_{\theta,T}$ by running $T$ iterations of the inner optimization dynamics $\Phi$, i.e., $w_{\theta,T} = \Phi(w_{\theta,T -1}, \theta) = \Phi(\Phi(w_{\theta,T-2}, \theta), \theta)$, and so on. In the general case where $\theta$ and $w$ are real-valued and the objectives and optimization dynamics is smooth, the gradient of the object $F (w_{\theta,T} )$ w.r.t. $\theta$, named hypergradient $\nabla_\theta F (w_{\theta,T} )$, can be computed by:
\begin{equation}
    \nabla_\theta F(w_{\theta,T})=\partial_w F(w_{\theta,T})\nabla_\theta w_{\theta,T},
\end{equation}
Here, the symbol $\partial$ represents the partial derivative used for computing the Jacobian with respect to immediate variables, whereas $\nabla$ signifies a total derivative that considers the recursive calls made to the function F. The gradient mentioned above can be efficiently computed by unfolding the optimization dynamics using back-propagation, which is essentially reverse mode automatic differentiation~\cite{griewank2008evaluating}. We repeatedly substitute $w_{\Phi,t} = \Phi(w_{\theta,t-1}, \theta)$ and apply the chain rule.

\section{Proposed Method}

 \begin{figure*}[h]
  \centering
  \includegraphics[scale=0.62]{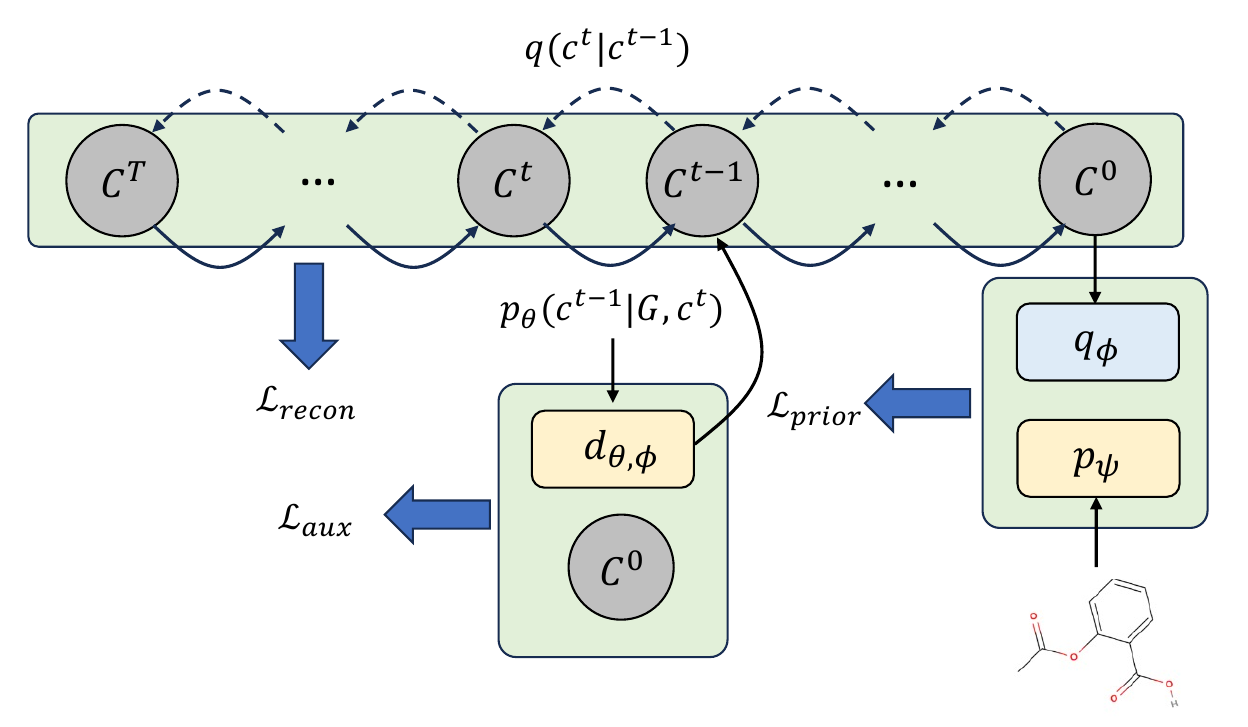}
  \caption{The overall framework of \method{}.}
  \label{framework}
\end{figure*}

In our framework, the authentic conformation $\mathbf{c}^0$ serves as a reference for a series of transformations through latent space, defined by $\mathbf{c}^t$ for $t = 1,\cdots, T$, where $t$ delineates the progression of diffusion steps. This is governed by a dual-phase probabilistic model~\cite{sohl2015deep}. Here, the initial phase disperses incremental noise through the conformational states, effectively transitioning from the true state to a noise-saturated latent space. Counteracting this, the subsequent phase engages in a systematic noise abatement, iteratively refining these perturbed states towards reconstructing the pristine conformation $\mathbf{c}^0$. A conceptual diagram outlining these mechanisms is depicted in Figure\ref{framework}.
 

\subsection{Diffusion Process} 
Drawing inspiration from thermodynamics, we conceptualize the conformations of molecules, denoted as $\textbf{C}$, as particles within a constantly fluctuating system. Over time, the initial equilibrium state $\textbf{c}^0$ undergoes a stochastic journey, progressing through a series of increasingly disordered states $\textbf{c}^t$, until it reaches a state of maximal entropy resembling white noise after $T$ transitional steps. This unique progression, distinct from conventional latent variable constructs, is a defining feature of our diffusion model. It is characterized by a predetermined posterior distribution, denoted as $q(\mathbf{c}^{1:T} |\mathbf{c}^0)$, which remains constant throughout the learning process. Specifically, we define this distribution as a Markov chain governed by a fixed variance schedule $\beta_1, \cdots , \beta_T$:
\begin{equation}
\begin{aligned}
    q(\mathbf{c}^{1:T} |\mathbf{c}^0)&=\Pi_{t=1}^T q(\mathbf{c}^t|\mathbf{c}^{t-1}),\\ q(\mathbf{c}^t|\mathbf{c}^{t-1})=\mathcal{N}&(\mathbf{c}^t;\sqrt{1-\beta_t}\mathbf{c}^{t-1},\beta_tI).
\end{aligned}
\end{equation}

It's important to note that in this work, we do not enforce specific invariance requirements on the diffusion process. The primary criterion is its efficiency in generating noisy samples for the training of the generative process, denoted as $p_\theta (\mathbf{c}^0)$.

By defining $\alpha_t = 1 - \beta_t$ and ${\bar{\alpha}}_t = \Pi^t_{s=1} \alpha_s$, we can observe a significant aspect of the forward process: the conditional distribution $q(\mathbf{c}^t|\mathbf{c}^0)$ at any time step $t$ can be precisely expressed as $q(\mathbf{c}^t|\mathbf{c}^0) = \mathcal{N} (\mathbf{c}^t; \sqrt{{\bar{\alpha}}_t} \mathbf{c}^0, (1-{\bar{\alpha}}_t)I)$. When we extend this process adequately up to $T$ steps, it inherently transforms the initial state $\mathbf{c}^0$ into a standard Gaussian distribution. This implies that a suitable choice for the prior distribution $p(\mathbf{c}^T)$ would be an isotropic Gaussian distribution.
 

\subsection{Reverse Process}
Our goal is to deduce the initial molecular conformations, denoted as $\mathbf{c}^0$, starting from a state of maximum entropy represented by white noise $\mathbf{c}^T$. This process is conditioned on the given molecular graph structures denoted as $G$. We approach this inference task as the temporal reversal of the diffusion sequence, commencing from the high-entropy state where the conformational coordinates $\mathbf{c}^T$ follow the distribution $p(\mathbf{c}^T)$. We formulate this reverse dynamics as a conditional Markov chain with learnable transitions:
\begin{equation}
\begin{aligned}
    &p_\theta(\mathbf{c}^{0:T-1}|G,\mathbf{c}^T)=\Pi_{t=1}^T p_\theta(\mathbf{c}^{t-1}|G,\mathbf{c}^t),\\
    p_\theta&(\mathbf{c}^{t-1}|G,\mathbf{c}^t)=\mathcal{N}(\mathbf{c}^{t-1};\mu_\theta(G,\mathbf{c}^t,t),\sigma_t^2I).
\end{aligned}
\end{equation}
In our framework, we utilize neural networks parameterized by $\theta$ to approximate mean values denoted as $\mu_\theta$, while $\sigma_t$ represents a variance that can be chosen by the practitioner. The process begins with the assignment of the initial state distribution $p(\mathbf{c}^T)$ as a standard Gaussian distribution. For any given molecular graph $G$, its three-dimensional configuration is constructed by initially sampling disordered state particles $\mathbf{c}^T$ from $p(\mathbf{c}^T)$. These particles are then progressively molded into the desired structure through a sequence of reverse Markov transitions $p_\theta (\mathbf{c}^{t-1}|G, \mathbf{c}^t)$.
 
Once we have established the mechanics of the inverse transformation, we calculate the conditional probability $p_\theta(\mathbf{c}^0|G)$ through an integration of the product of the initial state distribution and the sequential reverse dynamics. This is formally expressed as
$p_\theta(\mathbf{c}^0|G) =\int p(\mathbf{c}^T) p_\theta (\mathbf{c}^{0:T-1}|G, \mathbf{c}^T )d\mathbf{c}^{1:T} $. A crucial aspect of this computation is ensuring that our model's predictions remain consistent under any translation or rotation of the 3D structure. This concept is identified as vital for the generative process of three-dimensional forms. The following sections will elaborate on our approach to configuring the Markov transitions $p_\theta (\mathbf{c}^{t-1}|G, \mathbf{c}^t)$ to maintain this invariance. We will also discuss our strategy for optimizing the likelihood while adhering to these constraints.

\subsection{Training Objective}
To account for the multiplicity of stable molecular structures, our study employs a conditional variational autoencoder (CVAE) framework to construct the probabilistic model $p(\mathbf{C}|G)$. Within this framework, we introduce a latent variable $z$ to capture the intrinsic uncertainties associated with the generation of molecular conformations.
The CVAE architecture consists of a prior latent distribution $p_\psi (z|G)$ and a decoding mechanism $p_\theta (\mathbf{C}|z, G)$, which defines the conditional likelihood of the conformation $\mathbf{C}$ given $z$. Simultaneously, the training process incorporates an inferential component (encoder) $q_\psi(z|\mathbf{C}, G)$, enhancing the model's ability to learn from data. The encoder and decoder are jointly trained to maximize the evidence lower bound (ELBO) of the data log-likelihood:
\begin{equation}
\label{elbo}
\begin{aligned}
    \mathbb{E}&[log p_\theta (\mathbf{c}^0|G)]
    \geq\mathbb{E}\left[log \mathbb{E}_{q(\mathbf{c}^{1:T}|\mathbf{c}^0)} \frac{p_\theta(\mathbf{c}^{0:T}|G)}{q(\mathbf{c}^{1:T}|\mathbf{c}^0)}\right]\\
    &\quad-D_{KL}[q_\phi(z|\mathbf{C},G)||p_\psi(z|G)]\\
    &:=-\mathcal{L}_{ELBO}.
\end{aligned}
\end{equation}
This framework allows us to probabilistically model molecular conformations while accounting for inherent uncertainties and dependencies within the data.
The Evidence Lower Bound (ELBO), central to our model's training regimen, consists of two distinct components: the negative reconstruction loss, denoted as $L_{recon}$, and the latent space regularization term, represented as $L_{prior}$. The former assesses how well the reconstructed data matches the original data, while the latter promotes the maintenance of a well-behaved latent variable distribution, approximating the prior.
To capture the probabilistic nature of the model, both the output of the encoder, $q_\phi(z|\mathbf{C}, G)$, and the prior distribution, $p_\psi(z|G)$, are defined as diagonal Gaussian distributions, with parameters expressed as $\mathcal{N}(z|\mu_\phi(\mathbf{C}, G), \sigma_\phi(\mathbf{C}, G))$ and $\mathcal{N}(z|\mu_\psi(G), \sigma_\psi(G))$, respectively. The means and standard deviations of these distributions are determined by specialized graph neural networks.
During the optimization of the ELBO, we employ a reparametrization trick to enable backpropagation through the stochastic sampling process. Specifically, the latent variable $z$ is sampled by first computing $\mu_\phi(\mathbf{C}, G)$ and $\sigma_\phi(\mathbf{C}, G)$, and then adding a scaled random noise component $\epsilon$, where $\epsilon$ follows a standard Gaussian distribution $\mathcal{N}(0, I)$. This methodology facilitates gradient-based optimization by rendering the sampling process differentiable.

\subsection{Inner Objective}
To mitigate the challenges associated with rotational and translational dependencies in the generation of Cartesian coordinates for molecular conformations, contemporary methods often adopt an alternative approach. These techniques employ a decoder architecture specifically designed to generate inter-atomic distance matrices, denoted as $d_{\theta,\phi} = D_\theta(z_\phi, G)$, where these decoded distances are conditioned on the latent variable $z$ and the graph $G$. These distance matrices serve as a representation of the molecule's geometry that is invariant to translation and rotation.
Once the distance matrix $d$ is generated, it serves as the foundation for reconstructing the three-dimensional conformation $\mathbf{C}$. This reconstruction process involves solving a distance geometry problem, which essentially entails finding a set of coordinates that best align with the provided inter-atomic distances. The optimization process for this problem is often intricate and requires sophisticated techniques to determine the coordinates in a manner that accurately reflects the distances while adhering to physical plausibility.

\begin{equation}
\label{inner}
\begin{aligned}
    R_{\theta,\phi}&=arg\min_\mathbf{C} H(\mathbf{C}, D_\theta(z_\phi,G))\\
    &=arg\min_\mathbf{C} H(\mathbf{C},d_{\theta,\phi})\\
    &=arg\min_\mathbf{C} \left[\sum_{e_{uv}\in E}(||r_u-r_v||_2-d_{uv})^2 \right].
\end{aligned}
\end{equation}

\subsection{Outer Objective}
Our primary objective is to enhance the generation of 3D molecular structures in a way that closely aligns them with the actual models, regardless of their orientation or positioning. To quantify the precision of this alignment, we utilize the post-alignment Root-Mean-Square Deviation (RMSD), a commonly used metric in the field. This metric is calculated after optimally aligning a generated conformation $\mathbf{C}$ with a reference conformation $\mathbf{C}^*$ using an alignment operation $A$ that minimizes the disparity between the two. The RMSD is defined as the square root of the average squared distance between the atoms of the aligned generated conformation and the reference conformation after the alignment operation has been applied.
\begin{equation}
    RMSD(\mathbf{C},\hat{\mathbf{C}})=(\frac{1}{n}\sum_{i=1}^n ||\mathbf{c}_i-\hat{\mathbf{c}}_i||^2)^{\frac{1}{2}}.
\end{equation}
where $n$ is the number of atoms. Then the reconstruction objective $L_{recon}$ can be written as:
\begin{equation}
\label{outer}
\begin{aligned}
    F(R_{\theta,\phi})&=logp_\theta(\mathbf{C}|z,G)\\
    &=-\sum_{i=1}^n \sum_{j=1}^3 (\mathbf{C}_{ij}-A(\mathbf{C},\mathbf{C}^*)_{ij})^2,
\end{aligned}
\end{equation}
which is the outer loop objective for computing the reconstruction loss and maximize the log-likelihood.

\subsection{Bilevel Program}
To tackle the intricacies of molecular conformation generation, our approach employs a nested optimization scheme, where we distinguish between two interrelated objectives: one at a macro-level and another at a micro-level. The macro-level objective (referred to as the outer problem in Equation~\ref{outer}) aims to capture the conditional distribution $p(\mathbf{C}|G)$ of molecular structures, encompassing the full spectrum of conformations that molecules can assume based on their underlying graph $G$. Simultaneously, the micro-level objective (referred to as the inner problem in Equation~\ref{inner}) is focused on determining the precise molecular conformation from a set of predicted interatomic distances. By taking the expectation over the latent variable $z$, the resulting bilevel program for calculating the reconstruction term $L_{reconin}$ in Equation~\ref{elbo} can be expressed as follows:
\begin{equation}
\label{eq9}
    \max_{\theta,\phi}\mathbb{E}_{z\sim q_\phi(z|\mathbf{C},G)}[F(\mathbf{C}_{\theta,\phi},\theta)]
\end{equation}
\begin{equation}
\label{eq10}
    such\ that\ \mathbf{C}_{\theta,\phi}=arg\min_\mathbf{C} H(\mathbf{C},D_\theta(z_\phi,G)).    
\end{equation}

To handle the complexity of the bilevel optimization in our model, we address the absence of closed-form solutions and intractable expectations by utilizing a variational inference approach. This, combined with the reparametrization trick, offers a tractable, gradient-based estimation of molecular conformations, a technique we elaborate on in the following sections.

\subsection{Generative Model}
Our generative model for molecular conformations relies on the use of message-passing neural networks (MPNNs)~\cite{gilmer2017neural}, a class of graph neural networks known for their exceptional representation learning capabilities for molecular structures. MPNNs operate by updating atom embeddings through the aggregation of neighboring node information within the graph structure $G$, ensuring invariance to graph isomorphism and enabling state-of-the-art molecular modeling~\cite{scarselli2008graph,bruna2013spectral,kipf2017semisupervised}.
We have employed an MPNN architecture similar to the one described in~\cite{simm2019generative} for the encoder $q_\phi(z|\mathbf{C}, G)$ and the prior $p_\psi(z|G)$. Recognizing the intensive memory requirements of bilevel optimization, our decoder $p_\theta (\mathbf{C}|z, G)$ is innovatively implemented using a continuous normalizing flow (CNF) based on an ordinary differential equation (ODE), offering the advantage of a fixed memory footprint regardless of the number of layers. Details of our decoder's design are provided in the following sections.
In our architectural design, the decoder operates through two sequential stages. The initial stage involves a distance prediction network $D_\theta (z, G)$, which translates the latent space variable $z$ into an array of interatomic distances denoted by $d$. Subsequently, the model utilizes a differential geometry algorithm capable of reconstructing the molecular conformation $\mathbf{C}$ from the set of distances $d$. The network $D_\theta (z, G)$ is a conditional adaptation of the continuous normalizing flow (CNF) model. It begins with a noise distribution $d(t_0)$, representing the initial distances in the CNF ordinary differential equation (ODE) path, originating from a standard Gaussian distribution $\mathcal{N} (0, I)$. The path concludes with the final distances $d = d(t_1)$. This transformation is conditioned on the latent variable $z$ as well as the graph $G$:
\begin{equation}
\begin{aligned}
    d&=D_\theta(z,G)\\
    &=d(t_0)+\int_{t_0}^{t_1} g_\theta(d(t),t,G,z)dt,
\end{aligned}
\end{equation}
where $g_\theta$ is an MPNN that defines the continuous-time dynamics of the flow $D_\theta$ conditioned on $z$ and $G$. Note that, given the true distances $d(t_1) = d$, $d(t_0)$ can also be easily computed by reversing the continuous dynamics $D_\theta :D_\theta^{-1} (z, G) = d(t_1) + \int_{t_0}^{t_1} g_\theta (d(t), t, z, G)dt$. And thus the exact conditional log-likelihood of distances given $G$ can be computed by:
\begin{equation}
\begin{aligned}
    \mathcal{L}_{aux}&=log p_\theta(d|z,G)\\
    &=log p(d(t_0))-\int_{t_0}^{t_1}Tr(\frac{\partial g_\theta}{\partial d(t)})dt.
\end{aligned}
\end{equation}
In the training of our molecular conformation generative model, we utilize an ODE solver to integrate the continuous-time dynamics for backpropagation. This is complemented by an auxiliary objective $L_{aux}$, which imposes regularization on distance predictions. The integration process ensures that the model's parameters are optimized to produce accurate molecular geometries. It leverages the gradient estimations from the ODE solver and the guidance provided by $L_{aux}$ to enhance the accuracy of distance predictions, a crucial aspect of generating realistic molecular conformations. In summary, the training objective can be considered as the sum of three components:
\begin{equation}
    \mathcal{L}(\theta,\phi,\psi)=\mathcal{L}_{recon}+\lambda\mathcal{L}_{prior}+\alpha \mathcal{L}_{aux},
\end{equation}
where $\lambda$ and $\alpha$ are hyperparameters.

\subsection{Bilevel Optimization}
To address the optimization of the bilevel problem described in equation~\ref{eq9} and equation~\ref{eq10}, we propose a practical algorithm that builds upon existing methods used for inferring 3D structures from interatomic distances. The inner problem, as expressed in equation~\ref{eq10}, is analogous to traditional distance geometry challenges commonly encountered in molecular structure determination. Similar problems have been tackled using techniques such as semi-definite programming (SDP) for deducing protein configurations from NMR data and employing the Alternating Direction Method of Multipliers (ADMM) to computationally fold proteins into their native three-dimensional forms. Taking inspiration from these methodologies, our algorithm is designed to effectively address the intricacies of the inner optimization problem, facilitating the accurate reconstruction of molecular geometries from predicted interatomic distances. This approach enables the efficient generation of 3D coordinates that align with the provided pairwise distances, a critical aspect for the successful application of our model in predicting and generating molecular conformations.

In this initial work we choose gradient descent (GD), with tractable learning dynamics $\Phi$, to approximately solve for the geometry:
\begin{equation}
    \mathbf{C}_{\theta,\phi,t+1}=\Phi(\mathbf{C}_{\theta,\phi,t},d_{\theta,\phi})=\mathbf{C}_{\theta,\phi,t}-\eta \nabla H(\mathbf{C}_{\theta,\phi,t},d_{\theta,\phi}),
\end{equation}
where $\eta$ is the learning rate and $d_{\theta,\phi}$ is the distance set generated from the distance prediction model. Under appropriate assumptions and for a number of updates $t \to \infty$, GD can converge to a proper geometry $\mathbf{C}_{\theta,\phi}$ that depends on the predicted pairwise distances.

During the model training process, our focus is on calculating the hypergradient, which involves computing the derivative of the expected reconstruction error with respect to the model parameters. This derivative guides the adjustment of the model to improve conformation predictions. To compute the hypergradient, we generate molecular conformations through a series of gradient descent steps denoted as $\mathbf{C}_{\theta,\phi,T}$, which approximates the solution for the distance geometry problem.

Now we can write the hypergradient as:
\begin{equation}
\begin{aligned}
    \nabla_{\theta,\phi}&\mathbb{E}_{z\sim q_\phi(z|\mathbf{C},G)} [F (\mathbf{C}_{\theta,\phi,T})]\\
    &=\mathbb{E}_{z\sim q_\phi(z|\mathbf{C},G)}\partial_\mathbf{C} [F(\mathbf{C}_{\theta,\phi,T})] \nabla_{\theta,\phi}\mathbf{C}_{\theta,\phi,T}.
\end{aligned}
\end{equation}
To compute the gradient \(\nabla_{\theta,\phi}\mathbf{C}_{\theta,\phi,T}\), we meticulously trace the entire evolution of the inner optimization process, starting from the end state \(\mathbf{c}_T\) and working our way back to the initial state \(\mathbf{c}_0\). During the forward pass, we store each intermediate geometry \(\mathbf{c}_{0,\cdots,T}\) that arises from the inner optimization dynamics. When performing the backward computation, these saved geometries play a crucial role in calculating gradients through a sequence of Vector-Jacobian Products (VJPs). During the reverse computation, we propagate the gradient starting from \(\partial_{\mathbf{c}_T} F\) to the intermediate geometries \(\mathbf{c}_t\) using \(\nabla_{\mathbf{c}_t} \mathbf{c}_{t+1}\):
\begin{equation}
\label{eq16}
\begin{aligned}
        \nabla_{\mathbf{c}^t} \mathbf{c}^{t+1}&=\nabla_{\mathbf{c}^t}(\mathbf{c}^t-\eta \nabla_{\mathbf{c}^t}H(d_{\theta,\phi},\mathbf{c}^t))\\
        &=1-\eta \nabla_{\mathbf{c}^t}^2 H(d_{\theta,\phi},\mathbf{c}^t).
\end{aligned}
\end{equation}
The notation $\nabla_{\mathbf{c}^t}^2$ represents the Hessian matrix with respect to the conformation $\mathbf{c}^t$. By iteratively calculating the derivatives $\nabla_{\mathbf{c}^t} \mathbf{c}^{T}$, we can determine the adjoint values for the distances $d_{\theta,\phi}$. These adjoints are then used to perform Vector-Jacobian Products (VJPs), allowing the gradients to be propagated back through the network, affecting the parameters of the encoder $q_\phi$and the decoder $p_\theta$.
Formally, $\nabla_d \mathbf{c}^T$ is computed by:
\begin{equation}
\begin{aligned}
    \nabla_{d_{\theta,\psi}}\mathbf{c}^T&=\sum_{t=T-1}^0 [\nabla_{\mathbf{c}^{t+1}}\mathbf{c}^T]\nabla_d \mathbf{c}^{t+1}\\
    &=-\eta \sum_{t=T-1}^0 [\nabla_{\mathbf{c}^{t+1}}\mathbf{c}^T]\nabla_d(\nabla_{\mathbf{c}^t}H(d_{\theta,\phi},\mathbf{c}^t)),
\end{aligned}
\end{equation}
where $\nabla_{\mathbf{c}^{t+1}}\mathbf{c}^T$ can be substituted by equation~\ref{eq16}. The computation can be done efficiently with reverse-mode automatic differentiation software such as PyTorch. 

\begin{algorithm}[t]
\caption{Training Algorithm of \method{}}
\label{alg1}
\begin{algorithmic}[1]
\REQUIRE objective reweighting coefficients $\alpha$ and $\lambda$; the inner loop optimization iterations $T$ and learning rate $\eta$; alignment function $A(·, ·)$; data samples $\{G_t, \mathbf{c}^*_t \}$. \
\ENSURE prior $p_\psi (z|G)$, decoder $p_\theta (\mathbf{C}|z, G)$ and its dynamics defined as $g_\theta$ , encoder $q_\phi(z|\mathbf{C}, G)$.
\STATE Sample an initial molecular conformation $\mathbf{c}^T$ from a standard Gaussian distribution $\mathcal{N}(0, I)$
\FOR{$s=T,T-1,\cdots,1$}
\STATE Calculate $\mu_\theta(\mathbf{c}^s,G,s)$ from $\epsilon_\theta(\mathbf{c}^s,G,s)$;
\STATE Sample the next conformation $\mathbf{c}^{s-1}$ from a Gaussian distribution with mean $\mu_\theta(\mathbf{c}^s,G,s)$ and variance $\sigma^2_tI$
\STATE Compute $\mu_q$ and $\sigma_q$ and calculate the $\mathcal{L}_{prior}$ term.
\STATE $d^*\leftarrow R_t^*$;
\STATE Initialize the distance variables as $d^0$ and sample the initial distances $d(t_0)$ from a Gaussian distribution;
\STATE $d=D_\theta(z,G)=d(t_0)+\int_{t_0}^{t_1}g_\theta(d(t),t,G,z)dt$;
\FOR{$t=1,2,\cdots,T$}
\STATE $\mathbf{c}^{t+1}=\mathbf{c}^t-\eta \nabla H(\mathbf{c}^t,d)$;
\ENDFOR
\STATE $R \leftarrow \mathbf{c}^T$;
\STATE Calculate the reconstruction loss $\mathcal{L}_{recon}$;
\STATE Calculate  the total loss $\mathcal{L}$;
\STATE Update $\theta,\phi,\psi$;
\ENDFOR
\RETURN $q_\phi, p_\theta, p_\psi$

\end{algorithmic}
\end{algorithm}

\subsection{Sampling}
To generate a molecular conformation $\mathbf{C}$ based on a given molecular graph $G$, our process unfolds as follows: 
Firstly, we begin by sampling a latent variable $\Tilde{z}$ from the learned prior distribution $p_\psi(z|G)$.
Next, we acquire an initial set of interatomic distances $d(t_0)$ by sampling from a Gaussian distribution. These sampled distances, denoted as $\Tilde{d}(t_0)$, are a starting point for the distance prediction.
These sampled distances $\Tilde{d}(t_0)$ undergo processing through the trained Neural ODE $G_\theta$. This network, conditioned on $\Tilde{z}$ and $G$, transforms the initial distances $\Tilde{d}(t_0)$ into a new set of distances, denoted as $\Tilde{d}$.
Finally, we utilize these predicted distances $\Tilde{d}$ to reconstruct the molecular conformation $\mathbf{C}$. This is achieved by solving the optimization problem $\text{arg min}_\mathbf{C} H(\mathbf{C}, d_{\theta,\phi})$, which seeks to find the conformation $R$ that best matches the predicted distances, as defined in the previously mentioned equation.

\section{Experiments}
In the following sections, we will subject \method{} to comprehensive empirical testing, with a particular focus on its effectiveness in generating equilibrium conformations for molecules of various sizes, including those of significance in the field of pharmacology. To ensure a thorough evaluation, we will follow established practices in the field \cite{shi2021learning,ganea2021geomol} and compare our approach with other prominent contenders through two benchmark tests.
The first benchmark, known as $\textbf{Conformation Generation}$, will assess the model's capability to capture the conformational distribution. This evaluation will consider both the diversity and the fidelity of the conformations generated by the model.
The second benchmark, referred to as $\textbf{Property Prediction}$, will examine the model's accuracy in predicting molecular properties.
Before delving into the specific evaluation criteria for each task and discussing our findings, we will provide an overview of the general setup for our experiments.

\subsection{Experimental Settings}
\subsubsection{Datasets}
Our analysis is based on data from two primary datasets: GEOM-QM9 \cite{ramakrishnan2014quantum} and GEOM-Drugs \cite{ganea2021geomol}. These datasets encompass a wide range of molecular structures, with GEOM-QM9 primarily containing small molecules and GEOM-Drugs including medium-sized organic compounds. We have adopted the dataset partitions as defined by \cite{shi2021learning}.
Here is an overview of the dataset sizes and partitions:
\begin{itemize}
    \item  Training splits for both datasets consist of 40,000 molecules, each represented by 5 distinct conformations, resulting in a total of 200,000 conformations for training.
    \item  Validation sets are of the same size as the training splits.
    \item  The test split consists of 200 unique molecules, comprising 22,408 conformations for the QM9 dataset and 14,324 for the Drugs dataset.
\end{itemize}
This extensive collection of structural data provides a robust foundation for our comprehensive model evaluations.

The GEOM-QM9 dataset is an extension of the original QM9 dataset, enriched with multiple conformations for each molecule, primarily focusing on smaller compounds with a maximum of 9 heavy atoms. Our training data was obtained by extracting 50,000 pairs of conformations and molecules from this dataset. For testing purposes, we reserved 17,813 conformations derived from 150 molecular structures.
In contrast, the GEOM-Drugs dataset contains larger molecules characterized by an average of 6.5 rotatable bonds, reflecting complex molecular dynamics. From this dataset, we constructed our training set using an additional 50,000 pairs of conformations and molecules. For evaluation, we set aside 9,161 conformations originating from 100 molecular structures. This setup enables us to assess the performance of our model across a broad spectrum of molecular complexities.

\subsubsection{Evaluation metrics}
Evaluating the performance of conformation generation models involves assessing their ability to produce a diverse set of accurate molecular structures. The root-mean-square deviation (RMSD) metric is a crucial measure in this context, which quantifies the structural differences between the generated conformations and a reference set after aligning them using algorithms like Kabsch.
To comprehensively evaluate the models, several metrics are employed, including:
\begin{itemize}
    \item  Coverage: This metric measures the diversity of the generated conformations by indicating the proportion of reference conformations that are matched within a specified RMSD threshold.

    \item  Matching: Matching reflects the precision of the model by quantifying how many generated conformations closely resemble those in the reference set within a defined RMSD threshold.

    \item  Minimum RMSD: This metric provides insight into the average best-case alignment between the generated conformations and the reference set, indicating the overall accuracy of the model.

    \item  Maximum RMSD: The maximum RMSD highlights the worst outliers among the generated conformations, revealing cases where the model may struggle to produce accurate structures.
\end{itemize}
These metrics collectively offer a comprehensive assessment of the model's ability to generate diverse and accurate molecular conformations.
\begin{equation}
    COV-R(S_g, S_r)=\frac{1}{|S_r|}\left|\{\mathbf{c}\in S_r|RMSD(\mathbf{c},\hat{\mathbf{c}})\leq \delta, \hat{\mathbf{c}}\in S_g\}\right|,\nonumber
\end{equation}
\begin{equation}
    MAT-R(S_g,S_r)=\frac{1}{|S_r|}\sum_{\mathbf{c}\in S_r}\min_{\hat{\mathbf{c}}\in S_g} RMSD(\mathbf{c},\hat{\mathbf{c}}),
\end{equation}
The COV-P and MAT-P metrics provide a different perspective on evaluating conformation generation models by reversing the roles of the generated and reference sets. These metrics assess the performance of the models by doubling the size of the generated set $S_g$ relative to the reference set $S_r$ for each molecule. Here's how they work:
\begin{itemize}
    \item COV-P (Coverage-Precision): COV-P rates represent the percentage of reference structures that are adequately represented within the generated set based on a pre-defined RMSD threshold $\delta$. It measures how well the generated set covers the reference structures within the specified RMSD threshold. High COV-P rates indicate a good coverage of reference structures by the generated set.

    \item MAT-P (Mean RMSD-Precision): MAT-P scores reflect the mean RMSD between each generated conformation and its nearest reference counterpart. It calculates the average structural deviation between the generated and reference conformations. Low MAT-P scores indicate that the generated conformations closely resemble the reference structures in terms of structural similarity. 
\end{itemize}
Indeed, these metrics, including Precision, Recall, COV-P, and MAT-P, offer a well-rounded evaluation of a conformation generation model's performance. The selection of the RMSD threshold $\delta$ is crucial, as it determines what level of structural similarity is considered acceptable. Setting it to 0.5 Ångströms for the QM9 dataset and 1.25 Ångströms for the Drugs dataset is a common practice and aligns with established standards in the field. These threshold values are chosen to balance the need for accuracy and diversity in generated molecular structures, taking into account the specific characteristics of each dataset and the complexity of the molecules involved.

The property prediction task is essential for evaluating how well a generative model captures molecular properties across a range of generated conformations. It provides a direct assessment of the quality of the generated samples. In this task, a subset of 30 molecules from the GEOM-QM9 dataset is used to generate 50 samples for each molecule. The properties of interest, such as energy ($E$) and the HOMO-LUMO gap ($\epsilon$), are calculated for each generated conformation using the PSI4 chemical computation toolkit.
The following statistics are then compared against ground truth values to assess the model's performance:
\begin{itemize}
    \item  The average energy ($\Bar{E}$);
    \item The lowest energy conformer ($E_{min}$);
    \item  The average HOMO-LUMO gap ($\Bar{\Delta \epsilon}$);
    \item  The smallest gap ($\Delta \epsilon_{min}$);
    \item  The largest gap ($\Delta \epsilon_{max}$).
\end{itemize}
Discrepancies between the predicted values and ground truth values for these properties serve as indicators of the generative model's accuracy in reflecting true molecular properties. This evaluation provides valuable insights into how well the model captures the energetics and electronic properties of generated molecular conformations.

\subsection{Baselines}
The evaluation of our method, denoted as \method{}, is against four recent or established state-of-the-art baselines, inclusive of machine learning approaches and classic computational methods:
\begin{itemize}
    \item CVGAE~\cite{mansimov2019molecular}: This is a conditional Variational Autoencoder (VAE)-based model that employs several layers of graph neural networks to encode the atom representation from a molecular graph. It then directly predicts the 3D coordinates of molecules.

    \item GRAPHDG~\cite{simm2019generative}: Also utilizing a conditional VAE framework, this model differs from CVGAE in that it does not generate conformations directly in 3D space. Instead, it learns the distribution over interatomic distances and then employs a distance geometry algorithm to convert these distances into molecular conformations.

    \item CGCF~\cite{xu2021learning}: Standing for Continuous Geometric Conformational Flow, this two-stage method predicts atomic pairwise distances using continuous normalizing flows. The distances are then utilized to generate the conformations.

    \item RDKIT~\cite{Riniker_Landrum_2015}: A classical and widely utilized approach to conformation generation, RDKIT is an open-source toolkit that leverages a traditional distance geometry method with a comprehensive calculation of edge lengths derived from computational chemistry.
\end{itemize}

These baseline methods represent a spectrum from classical computational chemistry to modern machine learning models, providing a robust comparison across different strategies for molecular conformation generation. \method{} is tested against these baselines to determine its effectiveness and accuracy in generating realistic and diverse molecular conformations.

\subsection{Results and Analysis}

\begin{table}[t]
\centering
\tabcolsep=1.2pt
\caption{Comparison of different methods on the conformation generation task. }
\begin{tabular}{l|cccc|cccc}
\toprule
Dataset  &\multicolumn{4}{c}{GEOM-QM9}  &\multicolumn{4}{c}{GEOM-Drugs} \\
\multirow{2}*{Metric} &\multicolumn{2}{c}{COV$^*$ (\%)} &\multicolumn{2}{c}{MAT ($\mathring{A}$)} &\multicolumn{2}{c}{COV$^*$ (\%)} &\multicolumn{2}{c}{MAT ($\mathring{A}$)} \\
&Mean &Median &Mean &Median &Mean &Median &Mean &Median\\
\midrule 
CVGAE &8.52 &5.62 &0.7810 &0.7811 &0.00 &0.00 &2.5225 &2.4680\\
GraphDG &55.09 &56.47 &0.4649 &0.4298 &7.76 &0.00 &1.9840 &2.0108\\
CGCF &69.60 &70.64 &0.3915 &0.3986 &49.92 &41.07 &1.2698 &1.3064\\ 
ConfVAE &75.57 &80.76 &0.3873 &0.3850 &51.24 &46.36 &1.2487 &1.2609 \\
\method{} &\textbf{77.33} &\textbf{82.37} &\textbf{0.3742} &\textbf{0.3729} &\textbf{54.47} &\textbf{49.81} &\textbf{1.0254} &\textbf{1.0344}\\
\midrule 
\midrule 
RDKit &79.94 &87.20 &0.3238 &0.3195 &65.43 &70.00 &1.0962 &1.0877 \\ 
CVGAE+FF &63.10 &60.95 &0.3939 &0.4297 &83.08 &95.21 &0.9829 &0.9177\\
GraphDG+FF &70.67 &70.82 &0.4168 &0.3609 &84.68 &93.94 &0.9129 &0.9090\\ 
CGCF+FF &73.52 &72.75 &0.3131 &0.3251 &92.28 &98.15 &0.7740 &0.7338 \\
ConfVAE+FF &77.95 &79.14 &0.2851 &0.2817 &91.48 &99.21 &0.7743 &0.7436\\
\method{}+FF &\textbf{78.42} &\textbf{81.67} &\textbf{0.2745} &\textbf{0.2718} &\textbf{92.47} &\textbf{99.22} &\textbf{0.7716} &\textbf{0.7219}\\
\bottomrule
\end{tabular}
\label{tab::result}
\end{table}

The summarized results in Table~\ref{tab::result} provide a comprehensive performance assessment of molecular conformation generation methods. The table is divided into two sections, with the top five rows featuring deep generative models and the bottom six rows including various methods enhanced with an additional rule-based force field for optimizing the generated structures. The reported metrics include Coverage (COV) and Matching (MAT) scores, with Mean and Median values calculated across different molecular graphs within the GEOM test set. To ensure a fair and consistent evaluation, the size of the generated set is double that of the reference set, following the approach used in previous research~\cite{xu2021learning}.

The standout performance of \method{}, as reflected in the tables, indicates its exceptional capability to model the multimodal distribution of molecular conformations, delivering both precision and diversity. This is especially evident with the more complex and larger molecules found in the Drugs dataset, where \method{} surpasses state-of-the-art machine learning (ML) models by a significant margin. An interesting observation is the slight edge of \method{}-C over \method{}-A, hinting at the efficacy of the chain-rule approach in enhancing the optimization process. Given this finding, \method{}-C is selected for further comparisons.
These results demonstrate the effectiveness of \method{} in generating diverse and accurate molecular conformations, making it a promising tool for various applications in computational chemistry and drug discovery.
The performance of RDKIT on the intricate Drugs dataset was evaluated, and the results are presented in Table~\ref{tab::result}. These results show a consensus with previous studies that state-of-the-art machine learning (ML) models surpass RDKIT in Coverage-Recall (COV-R) and Matching-Recall (MAT-R) metrics. However, ML models still fall short when it comes to the newly introduced Precision-based metrics. This discrepancy suggests that while ML models are more exploratory in identifying a broader range of potential structures, RDKIT tends to focus on producing a smaller set of the most probable structures, emphasizing quality over quantity. This tendency of RDKIT to produce higher quality at the expense of diversity has been attributed to its use of an empirical force field (FF), a strategy employed in traditional computational chemistry to refine and optimize molecular structures.
In an attempt to level the playing field and conduct a more equitable comparison, \method{} was combined with an empirical force field to create the \method{} + FF model. The results show that the \method{} + FF model was able to maintain the superior diversity (as reflected by the Recall metrics) that ML models are known for, while also achieving a significant boost in accuracy (as evidenced by the Precision metrics). This outcome suggests that integrating \method{} with a force field is a potent strategy, ensuring that the model retains its capacity to generate a wide range of conformations (thereby capturing the complex multimodal nature of molecular structures), while also enhancing the precision and reliability of the conformations it predicts. It's an effective fusion of the exploratory power of ML with the targeted refinement capabilities of force fields.


\subsection{Property Prediction}
The mean absolute errors (MAE) reported in Table~\ref{property} for various properties of molecules generated by different models provide insights into how accurately a model can predict these properties when compared to ground truth data obtained from experimental or highly accurate quantum chemical calculations. The properties considered in this evaluation are highly sensitive to the geometric structure of molecules, meaning that even slight changes in the distances between atoms can lead to significant differences in the predicted properties. Therefore, a model with low MAE for these properties demonstrates its ability to consistently generate molecular conformations that closely resemble true molecular structures.
In summary, the low MAE reported for \method{} suggests that it not only generates conformations with high diversity but also with a high degree of geometric accuracy. This accuracy translates into effective predictions of properties that are closely aligned with the actual properties of the molecules. This reinforces the idea that \method{} is capable of capturing the complex underlying distribution of molecular structures in a way that is both diverse and precise.

\begin{table}[t]
\centering
\tabcolsep=5pt
\caption{MAE of predicted ensemble properties in eV. }
\begin{tabular}{l|ccccc}
\toprule
Method  &$\Bar{E}$  &$E_{min}$ &$\Bar{\Delta \epsilon}$ &$\Delta \epsilon_{min}$ &$\Delta \epsilon_{max}$\\
\midrule 
RDKit &0.9233 &0.6585 &0.3698 &0.8021 &0.2359\\
GraphDG &9.1027 &0.8882 &1.7973 &4.1743 &0.4776\\
CGCF &28.9661 &2.8410 &2.8356 &10.6361 &0.5954\\ 
ConfVAE &8.2080 &0.6100 &1.6080 &3.9111 &0.2429\\
\method{} &\textbf{0.8872} &\textbf{0.5962} &\textbf{0.3874} &\textbf{0.7782} &\textbf{0.2247}\\
\bottomrule
\end{tabular}
\label{property}
\end{table}
\section{Conclusion}
In conclusion, this study introduces a groundbreaking generative framework, \method{}, that represents a significant advancement in the accurate prediction of three-dimensional molecular structures from their two-dimensional graph representations. By drawing inspiration from non-equilibrium thermodynamics and leveraging diffusion principles, \method{} offers a cohesive approach to transforming stochastic distributions into precise molecular conformations. Its novel bilevel optimization process, designed to address the complexities of molecular geometry, and its commitment to maintaining roto-translational invariance set new standards in the field of computational chemistry. The extensive empirical evaluations presented in this study demonstrate \method{}'s superiority over existing state-of-the-art methods, establishing its significance as a vital tool for advancements in drug development and a deeper understanding of molecular dynamics and interactions.

\bibliographystyle{IEEEtran}
\bibliography{reference}

\begin{thebibliography}{10}
\providecommand{\url}[1]{#1}
\csname url@samestyle\endcsname
\providecommand{\newblock}{\relax}
\providecommand{\bibinfo}[2]{#2}
\providecommand{\BIBentrySTDinterwordspacing}{\spaceskip=0pt\relax}
\providecommand{\BIBentryALTinterwordstretchfactor}{4}
\providecommand{\BIBentryALTinterwordspacing}{\spaceskip=\fontdimen2\font plus
\BIBentryALTinterwordstretchfactor\fontdimen3\font minus \fontdimen4\font\relax}
\providecommand{\BIBforeignlanguage}[2]{{%
\expandafter\ifx\csname l@#1\endcsname\relax
\typeout{** WARNING: IEEEtran.bst: No hyphenation pattern has been}%
\typeout{** loaded for the language `#1'. Using the pattern for}%
\typeout{** the default language instead.}%
\else
\language=\csname l@#1\endcsname
\fi
#2}}
\providecommand{\BIBdecl}{\relax}
\BIBdecl

\bibitem{yin2022deal}
N.~Yin, L.~Shen, B.~Li, M.~Wang, X.~Luo, C.~Chen, Z.~Luo, and X.-S. Hua, ``Deal: An unsupervised domain adaptive framework for graph-level classification,'' in \emph{Proceedings of the 30th ACM International Conference on Multimedia}, 2022, pp. 3470--3479.

\bibitem{duvenaud2015convolutional}
D.~K. Duvenaud, D.~Maclaurin, J.~Iparraguirre, R.~Bombarell, T.~Hirzel, A.~Aspuru-Guzik, and R.~P. Adams, ``Convolutional networks on graphs for learning molecular fingerprints,'' in \emph{Proceedings of the Conference on Neural Information Processing Systems}, vol.~28, 2015.

\bibitem{anonymous2024dream}
N.~Yin, L.~Shen, H.~Xiong, B.~Gu, Z.~Chen, M.~Wang, and X.~Luo, ``{DREAM}: Dual structured exploration with mixup for open-set graph domain adaption,'' in \emph{Proceedings of the International Conference on Learning Representations}, 2024.

\bibitem{yin2023}
J.~Pang, Z.~Wang, J.~Tang, M.~Xiao, and N.~Yin, ``Sa-gda: Spectral augmentation for graph domain adaptation,'' in \emph{Proceedings of the 31st ACM International Conference on Multimedia}, 2023, pp. 309--318.

\bibitem{yin2023coco}
N.~Yin, L.~Shen, M.~Wang, L.~Lan, Z.~Ma, C.~Chen, X.-S. Hua, and X.~Luo, ``Coco: A coupled contrastive framework for unsupervised domain adaptive graph classification,'' in \emph{The 40th International Conference on Machine Learning}, 2023.

\bibitem{yin2022dynamic}
N.~Yin, F.~Feng, Z.~Luo, X.~Zhang, W.~Wang, X.~Luo, C.~Chen, and X.-S. Hua, ``Dynamic hypergraph convolutional network,'' in \emph{2022 IEEE 38th International Conference on Data Engineering (ICDE)}.\hskip 1em plus 0.5em minus 0.4em\relax IEEE, 2022, pp. 1621--1634.

\bibitem{yin2023messages}
N.~Yin, L.~Shen, H.~Xiong, B.~Gu, C.~Chen, X.-S. Hua, S.~Liu, and X.~Luo, ``Messages are never propagated alone: Collaborative hypergraph neural network for time-series forecasting,'' in \emph{IEEE Transactions on Pattern Analysis and Machine Intelligence}.\hskip 1em plus 0.5em minus 0.4em\relax IEEE, 2023.

\bibitem{Jin_Barzilay_Jaakkola_2017}
W.~Jin, R.~Barzilay, and T.~Jaakkola, ``Junction tree variational autoencoder for molecular graph generation,'' in \emph{Proceedings of the International Conference on Machine Learning}, 2017.

\bibitem{shi2020graphaf}
C.~Shi, M.~Xu, Z.~Zhu, W.~Zhang, M.~Zhang, and J.~Tang, ``Graphaf: a flow-based autoregressive model for molecular graph generation,'' in \emph{Proceedings of the International Conference on Learning Representations}, 2020.

\bibitem{thomas2018tensor}
N.~Thomas, T.~Smidt, S.~Kearnes, L.~Yang, L.~Li, K.~Kohlhoff, and P.~Riley, ``Tensor field networks: Rotation-and translation-equivariant neural networks for 3d point clouds,'' \emph{arXiv preprint arXiv:1802.08219}, 2018.

\bibitem{gebauer2022inverse}
N.~W. Gebauer, M.~Gastegger, S.~S. Hessmann, K.-R. M{\"u}ller, and K.~T. Sch{\"u}tt, ``Inverse design of 3d molecular structures with conditional generative neural networks,'' \emph{Nature communications}, vol.~13, no.~1, p. 973, 2022.

\bibitem{jing2021learning}
B.~Jing, S.~Eismann, P.~Suriana, R.~J.~L. Townshend, and R.~Dror, ``Learning from protein structure with geometric vector perceptrons,'' in \emph{Proceedings of the International Conference on Learning Representations}, 2021.

\bibitem{batzner20223}
S.~Batzner, A.~Musaelian, L.~Sun, M.~Geiger, J.~P. Mailoa, M.~Kornbluth, N.~Molinari, T.~E. Smidt, and B.~Kozinsky, ``E (3)-equivariant graph neural networks for data-efficient and accurate interatomic potentials,'' \emph{Nature communications}, vol.~13, no.~1, p. 2453, 2022.

\bibitem{Hawkins_2017}
\BIBentryALTinterwordspacing
P.~C.~D. Hawkins, ``\BIBforeignlanguage{en-US}{Conformation generation: The state of the art},'' \emph{\BIBforeignlanguage{en-US}{Journal of Chemical Information and Modeling}}, p. 1747–1756, Aug 2017. [Online]. Available: \url{http://dx.doi.org/10.1021/acs.jcim.7b00221}
\BIBentrySTDinterwordspacing

\bibitem{de2016role}
M.~De~Vivo, M.~Masetti, G.~Bottegoni, and A.~Cavalli, ``Role of molecular dynamics and related methods in drug discovery,'' \emph{Journal of medicinal chemistry}, vol.~59, no.~9, pp. 4035--4061, 2016.

\bibitem{ballard2015exploiting}
A.~J. Ballard, S.~Martiniani, J.~D. Stevenson, S.~Somani, and D.~J. Wales, ``Exploiting the potential energy landscape to sample free energy,'' \emph{Wiley Interdisciplinary Reviews: Computational Molecular Science}, vol.~5, no.~3, pp. 273--289, 2015.

\bibitem{mansimov2019molecular}
E.~Mansimov, O.~Mahmood, S.~Kang, and K.~Cho, ``Molecular geometry prediction using a deep generative graph neural network,'' \emph{Scientific reports}, vol.~9, no.~1, p. 20381, 2019.

\bibitem{simm2019generative}
G.~N. Simm and J.~M. Hern{\'a}ndez-Lobato, ``A generative model for molecular distance geometry,'' in \emph{Proceedings of the International Conference on Machine Learning}, 2020.

\bibitem{xu2021learning}
M.~Xu, S.~Luo, Y.~Bengio, J.~Peng, and J.~Tang, ``Learning neural generative dynamics for molecular conformation generation,'' \emph{arXiv preprint arXiv:2102.10240}, 2021.

\bibitem{Liberti_Lavor_Maculan_Mucherino_2012}
L.~Liberti, C.~Lavor, N.~Maculan, and A.~Mucherino, ``\BIBforeignlanguage{en-US}{Euclidean distance geometry and applications},'' \emph{\BIBforeignlanguage{en-US}{Siam Review,Siam Review}}, May 2012.

\bibitem{senior2020improved}
A.~W. Senior, R.~Evans, J.~Jumper, J.~Kirkpatrick, L.~Sifre, T.~Green, C.~Qin, A.~{\v{Z}}{\'\i}dek, A.~W. Nelson, A.~Bridgland \emph{et~al.}, ``Improved protein structure prediction using potentials from deep learning,'' \emph{Nature}, vol. 577, no. 7792, pp. 706--710, 2020.

\bibitem{jumper2020high}
J.~Jumper, R.~Evans, A.~Pritzel, T.~Green, M.~Figurnov, K.~Tunyasuvunakool, O.~Ronneberger, R.~Bates, A.~{\v{Z}}{\'\i}dek, A.~Bridgland \emph{et~al.}, ``High accuracy protein structure prediction using deep learning,'' \emph{Fourteenth critical assessment of techniques for protein structure prediction (abstract book)}, vol.~22, no.~24, p.~2, 2020.

\bibitem{franceschi2018bilevel}
L.~Franceschi, P.~Frasconi, S.~Salzo, R.~Grazzi, and M.~Pontil, ``Bilevel programming for hyperparameter optimization and meta-learning,'' in \emph{International conference on machine learning}.\hskip 1em plus 0.5em minus 0.4em\relax PMLR, 2018, pp. 1568--1577.

\bibitem{yin2022generic}
N.~Yin and Z.~Luo, ``Generic structure extraction with bi-level optimization for graph structure learning,'' \emph{Entropy}, vol.~24, no.~9, p. 1228, 2022.

\bibitem{kohler2020equivariant}
J.~K{\"o}hler, L.~Klein, and F.~No{\'e}, ``Equivariant flows: exact likelihood generative learning for symmetric densities,'' in \emph{Proceedings of the International Conference on Machine Learning}.\hskip 1em plus 0.5em minus 0.4em\relax PMLR, 2020, pp. 5361--5370.

\bibitem{xu2021end}
M.~Xu, W.~Wang, S.~Luo, C.~Shi, Y.~Bengio, R.~Gomez-Bombarelli, and J.~Tang, ``An end-to-end framework for molecular conformation generation via bilevel programming,'' in \emph{Proceedings of the International Conference on Machine Learning}.\hskip 1em plus 0.5em minus 0.4em\relax PMLR, 2021, pp. 11\,537--11\,547.

\bibitem{shi2021learning}
C.~Shi, S.~Luo, M.~Xu, and J.~Tang, ``Learning gradient fields for molecular conformation generation,'' in \emph{Proceedings of the International Conference on Machine Learning}.\hskip 1em plus 0.5em minus 0.4em\relax PMLR, 2021, pp. 9558--9568.

\bibitem{luo2021predicting}
S.~Luo, C.~Shi, M.~Xu, and J.~Tang, ``Predicting molecular conformation via dynamic graph score matching,'' in \emph{Proceedings of the Conference on Neural Information Processing Systems}, vol.~34, 2021, pp. 19\,784--19\,795.

\bibitem{song2019generative}
Y.~Song and S.~Ermon, ``Generative modeling by estimating gradients of the data distribution,'' in \emph{Proceedings of the Conference on Neural Information Processing Systems}, vol.~32, 2019.

\bibitem{song2020improved}
------, ``Improved techniques for training score-based generative models,'' in \emph{Proceedings of the Conference on Neural Information Processing Systems}, vol.~33, 2020.

\bibitem{hendrycks2016baseline}
D.~Hendrycks and K.~Gimpel, ``A baseline for detecting misclassified and out-of-distribution examples in neural networks,'' \emph{arXiv preprint arXiv:1610.02136}, 2016.

\bibitem{ganea2021geomol}
O.~Ganea, L.~Pattanaik, C.~Coley, R.~Barzilay, K.~Jensen, W.~Green, and T.~Jaakkola, ``Geomol: Torsional geometric generation of molecular 3d conformer ensembles,'' in \emph{Proceedings of the Conference on Neural Information Processing Systems}, vol.~34, 2021, pp. 13\,757--13\,769.

\bibitem{TorsionNet}
T.~Gogineni, Z.~Xu, E.~Punzalan, R.~Jiang, J.~Kammeraad, A.~Tewari, and P.~Zimmerman, ``Torsionnet: A reinforcement learning approach to sequential conformer search,'' in \emph{Proceedings of the Conference on Neural Information Processing Systems}, 2020.

\bibitem{noe2019boltzmann}
F.~No{\'e}, S.~Olsson, J.~K{\"o}hler, and H.~Wu, ``Boltzmann generators: Sampling equilibrium states of many-body systems with deep learning,'' \emph{Science}, vol. 365, no. 6457, p. eaaw1147, 2019.

\bibitem{alquraishi2019end}
M.~AlQuraishi, ``End-to-end differentiable learning of protein structure,'' \emph{Cell systems}, vol.~8, no.~4, pp. 292--301, 2019.

\bibitem{ingraham2018learning}
J.~Ingraham, A.~Riesselman, C.~Sander, and D.~Marks, ``Learning protein structure with a differentiable simulator,'' in \emph{Proceedings of the International Conference on Learning Representations}, 2019.

\bibitem{jumper2021highly}
J.~Jumper, R.~Evans, A.~Pritzel, T.~Green, M.~Figurnov, O.~Ronneberger, K.~Tunyasuvunakool, R.~Bates, A.~{\v{Z}}{\'\i}dek, A.~Potapenko \emph{et~al.}, ``Highly accurate protein structure prediction with alphafold,'' \emph{Nature}, vol. 596, no. 7873, pp. 583--589, 2021.

\bibitem{luo2021diffusion}
S.~Luo and W.~Hu, ``Diffusion probabilistic models for 3d point cloud generation,'' in \emph{Proceedings of the IEEE/CVF Conference on Computer Vision and Pattern Recognition}, 2021, pp. 2837--2845.

\bibitem{chibane2020implicit}
J.~Chibane, T.~Alldieck, and G.~Pons-Moll, ``Implicit functions in feature space for 3d shape reconstruction and completion,'' in \emph{Proceedings of the IEEE/CVF Conference on Computer Vision and Pattern Recognition}, 2020, pp. 6970--6981.

\bibitem{weiler20183d}
M.~Weiler, M.~Geiger, M.~Welling, W.~Boomsma, and T.~S. Cohen, ``3d steerable cnns: Learning rotationally equivariant features in volumetric data,'' in \emph{Proceedings of the Conference on Neural Information Processing Systems}, vol.~31, 2018.

\bibitem{fuchs2020se}
F.~Fuchs, D.~Worrall, V.~Fischer, and M.~Welling, ``Se (3)-transformers: 3d roto-translation equivariant attention networks,'' in \emph{Proceedings of the Conference on Neural Information Processing Systems}, vol.~33, 2020, pp. 1970--1981.

\bibitem{miller2020relevance}
B.~K. Miller, M.~Geiger, T.~E. Smidt, and F.~No{\'e}, ``Relevance of rotationally equivariant convolutions for predicting molecular properties,'' \emph{arXiv preprint arXiv:2008.08461}, 2020.

\bibitem{satorras2021n}
V.~G. Satorras, E.~Hoogeboom, and M.~Welling, ``E (n) equivariant graph neural networks,'' in \emph{Proceedings of the International Conference on Machine Learning}.\hskip 1em plus 0.5em minus 0.4em\relax PMLR, 2021, pp. 9323--9332.

\bibitem{garcia2021n}
V.~Garcia~Satorras, E.~Hoogeboom, F.~Fuchs, I.~Posner, and M.~Welling, ``E (n) equivariant normalizing flows,'' in \emph{Proceedings of the Conference on Neural Information Processing Systems}, vol.~34, 2021, pp. 4181--4192.

\bibitem{griewank2008evaluating}
A.~Griewank and A.~Walther, \emph{Evaluating derivatives: principles and techniques of algorithmic differentiation}.\hskip 1em plus 0.5em minus 0.4em\relax SIAM, 2008.

\bibitem{sohl2015deep}
J.~Sohl-Dickstein, E.~Weiss, N.~Maheswaranathan, and S.~Ganguli, ``Deep unsupervised learning using nonequilibrium thermodynamics,'' in \emph{Proceedings of the International Conference on Machine Learning}.\hskip 1em plus 0.5em minus 0.4em\relax PMLR, 2015, pp. 2256--2265.

\bibitem{gilmer2017neural}
J.~Gilmer, S.~S. Schoenholz, P.~F. Riley, O.~Vinyals, and G.~E. Dahl, ``Neural message passing for quantum chemistry,'' in \emph{Proceedings of the International Conference on Machine Learning}.\hskip 1em plus 0.5em minus 0.4em\relax PMLR, 2017, pp. 1263--1272.

\bibitem{scarselli2008graph}
F.~Scarselli, M.~Gori, A.~C. Tsoi, M.~Hagenbuchner, and G.~Monfardini, ``The graph neural network model,'' \emph{IEEE transactions on neural networks}, vol.~20, no.~1, pp. 61--80, 2008.

\bibitem{bruna2013spectral}
J.~Bruna, W.~Zaremba, A.~Szlam, and Y.~LeCun, ``Spectral networks and locally connected networks on graphs,'' \emph{arXiv preprint arXiv:1312.6203}, 2013.

\bibitem{kipf2017semisupervised}
T.~N. Kipf and M.~Welling, ``Semi-supervised classification with graph convolutional networks,'' in \emph{Proceedings of the International Conference on Learning Representations}, 2017.

\bibitem{ramakrishnan2014quantum}
R.~Ramakrishnan, P.~O. Dral, M.~Rupp, and O.~A. Von~Lilienfeld, ``Quantum chemistry structures and properties of 134 kilo molecules,'' \emph{Scientific data}, vol.~1, no.~1, pp. 1--7, 2014.

\bibitem{Riniker_Landrum_2015}
\BIBentryALTinterwordspacing
S.~Riniker and G.~A. Landrum, ``\BIBforeignlanguage{en-US}{Better informed distance geometry: Using what we know to improve conformation generation.}'' \emph{\BIBforeignlanguage{en-US}{Journal of Chemical Information and Modeling}}, p. 2562–2574, Dec 2015. [Online]. Available: \url{http://dx.doi.org/10.1021/acs.jcim.5b00654}
\BIBentrySTDinterwordspacing

\end{thebibliography}

\begin{IEEEbiography}[{\includegraphics[width=1in,height=1.25in,clip,keepaspectratio]{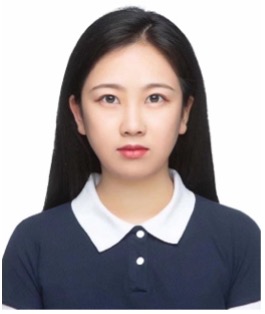}}]{Bobin Yang} is a dedicated student pursuing a bachor's degree at the University of Chinese Academy of Sciences. She completed her undergraduate studies at Northwestern University in China. Currently enrolled in the Life Sciences program at the University of Chinese Academy of Sciences, Bobin's primary research focus revolves around the application of machine learning and deep learning techniques in protein prediction, as well as the analysis of single-cell and spatial omics data.
\end{IEEEbiography}

\begin{IEEEbiography}[{\includegraphics[width=1in,height=1.25in,clip,keepaspectratio]{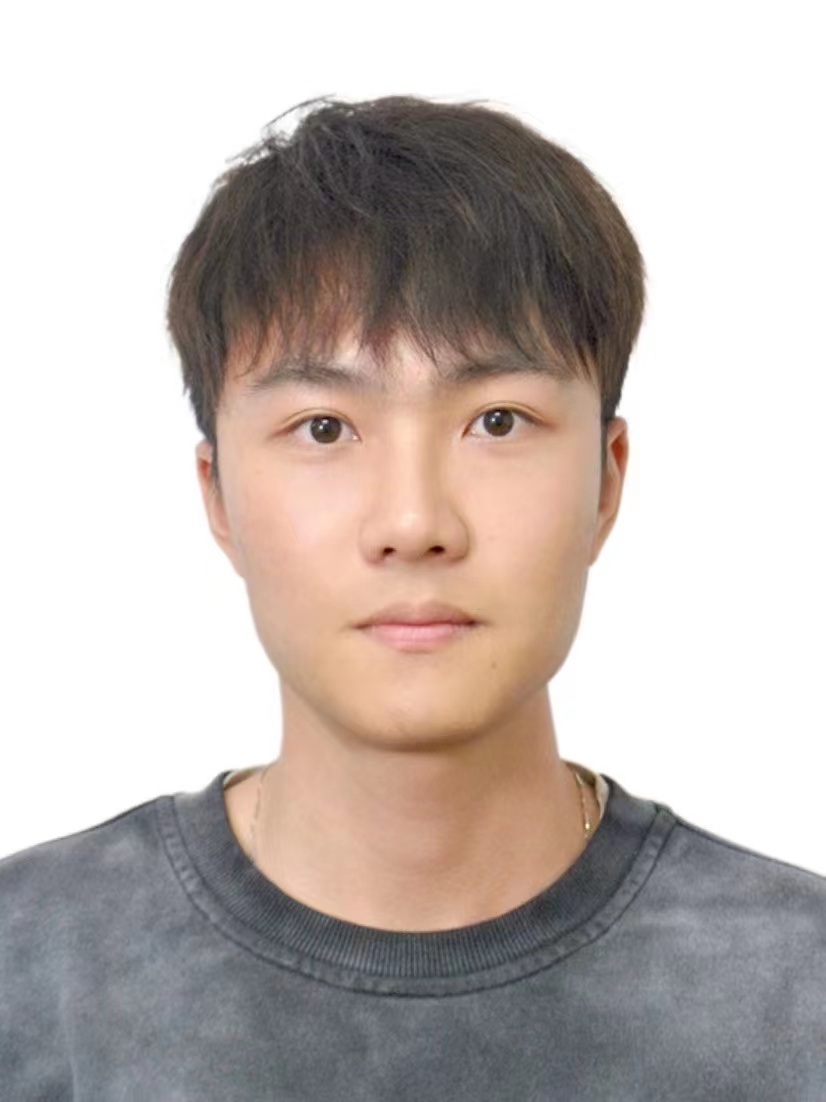}}]{Jie Deng} is a Software Engineering student at the Nankai University who responsible for the development of an App market based on Android for internal usage: built landing page, installed apps page and activities layouts using XML; implemented multiple functions: multi-thread downloading, resume from breakpoint, installation, etc. by using OKHTTP Library.
\end{IEEEbiography}

\begin{IEEEbiography}[{\includegraphics[width=1in,height=1.25in,clip,keepaspectratio]{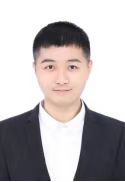}}]{Zhenghan Chen} is currently a Kaggle Master and  a applied scientist doing research on AI at Yanshi. His current research interest include LLM, AIGent and so on.
\end{IEEEbiography}

\begin{IEEEbiography}[{\includegraphics[width=1in,height=1.25in,clip,keepaspectratio]{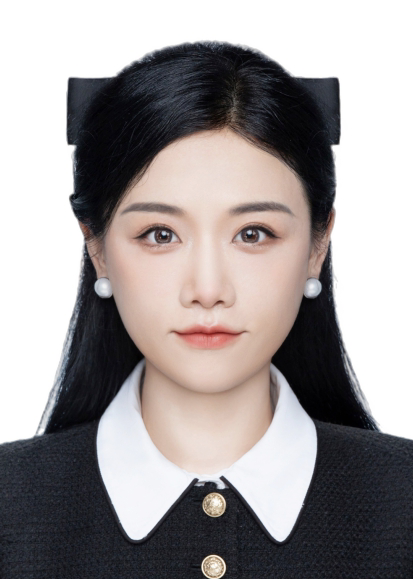}}]{Ruoxue Wu} received the B.E. degree in software engineering from Northwest University, China, and the M.S. degree in software engineering from Yunnan University, China. Her research interests include artificial intelligence, deep learning, and data visualization.
\end{IEEEbiography}

\end{document}